\documentclass[a4paper,useAMS,usenatbib]{mn2e}
\usepackage[totalwidth=480pt, totalheight=680pt]{geometry}
\usepackage{graphicx}

\def\gr{{$\gamma$-ray}}
\def\mw{{$\mu$-wave}}
\def\ergs{{erg~cm$^{-2}$~s$^{-1}$~}}

\def\lognulnu{Log$(\nu L \nu_{\rm 5GHz})$}
\def\nulnu{$\nu L \nu_{\rm 5GHz}$}
\newcommand{\lsim}{{\lower.5ex\hbox{$\; \buildrel < \over \sim \;$}}}
\newcommand{\gsim}{{\lower.5ex\hbox{$\; \buildrel > \over \sim \;$}}}
\def\nup{$\nu_{\rm peak}^S$}
 
\newdimen\digitwidth
\setbox0=\hbox{2}
\digitwidth=\wd0
\catcode `#=\active
\def#{\kern\digitwidth}

\newcommand{\fermi}{{\it Fermi}}

\newcommand{\paperone}{{Paper I}}

\newcommand{\paperthree}{{Paper III}}

\title[A simplified view of blazars: contribution to the X-ray and \gr\ cosmic backgrounds]{A simplified view of
  blazars: contribution to the X-ray and \gr\ extragalactic  backgrounds
 }
 \author[P. Giommi and P. Padovani]{ P. Giommi$^{1,2}$\thanks{E-mail:
paolo.giommi@asi.it}, P. Padovani$^{3,4}$\\
$^{1}$ASI Science Data Center, via del Politecnico s.n.c., I-00133 Roma Italy \\
$^{2}$ICRANet-Rio, CBPF, Rua Dr. Xavier Sigaud 150, 22290-180 Rio de Janeiro, Brazil\\
$^{3}$European Southern Observatory, Karl-Schwarzschild-Str. 2,
D-85748 Garching bei M\"unchen, Germany\\
$^{4}$Associated to INAF - Osservatorio Astronomico di Roma, via Frascati 33,
I-00040 Monteporzio Catone, Italy\\
}

\begin{document}

\date{Accepted ... Received ...; in original form ...}

\pagerange{\pageref{firstpage}--\pageref{lastpage}} \pubyear{2015}

\maketitle

\label{firstpage}

\begin{abstract}

The {\it blazar simplified view} is a new paradigm that explains well the diverse statistical properties of blazars observed over the entire electromagnetic spectrum on the basis of minimal assumptions on blazars' physical and geometrical properties.  
In this paper, the fourth in a series, we extend the predictions of this paradigm below the sensitivity of existing surveys and estimate the contribution of blazars to the 
X-ray and  \gr\  extragalactic backgrounds. We find that the integrated light from blazars can explain up to 100\% of the cosmic  background at energies larger than $\sim $10 GeV, and contribute $\approx 40 - 70\%$ of the \gr\ diffuse radiation between 100 MeV and 10 GeV. The contribution of blazars to the X-ray background, between 1 and 50 keV, is 
approximately constant and of the order of $4-5\%$.   On the basis of an interpolation between the estimated flux at X-ray and \gr\ energies we can expect that the contribution of blazars raises to $\sim$ 10\% at 100 keV, and continues to increase with energy until it becomes the dominant component at a few MeV.  Finally, 
we show that a strong dependence of the  synchrotron peak frequency on luminosity, as postulated by the {\it blazar sequence}, is ruled out by the observational data as it predicts a \gr\ background above a few GeV that is far in excess of the observed value.
%implications for AGN synthesis models...

  \end{abstract}


%\begin{document}
%-------------------------------------------------------------------------------

\section{Introduction}\label{intro}

The extragalactic sky from any direction and at all frequencies is filled with radiation from discrete sources and from a diffuse (or unresolved) component known as the Cosmic Background. This pervasive radiation, discovered only in relatively recent times, is one of the most fundamental observables from the Universe, as it carries crucial information on the integrated radiation emitted over the entire cosmic history. 
The first extragalactic background to be detected was the Cosmic X-ray Background \citep[CXB,][]{Giacconi62}, a few years before the discovery of the Cosmic Microwave Background \citep[CMB,][]{penziasandwilson65}, which is much brighter in terms of energy density, and is truly diffuse. The Extragalactic Gamma-Ray Background (EGB) was first detected by the SAS2 satellite in the 1970's  \citep{fichtel77}, and for a long time its nature was poorly understood. The latest Fermi-LAT survey data are now providing strong evidence towards an origin mostly due to integrated radiation from blazars and radio galaxies \citep[e.g.][]{inoue14,ajello2015,fornasa15}.
Blazars are a special type of extragalactic sources showing extreme observational properties, such as rapid and large amplitude variability, superluminal motion, and strong non-thermal emission across the entire electromagnetic spectrum. Theses sources are thought to be Active Galactic Nuclei (AGN) that host a jet pointing almost directly to the observer, within which relativistic particles move and radiate by losing their energy in a magnetic field \citep{bla78,UP95}. 
Among extragalactic objects blazars are known to accelerate particles to the highest observed energies and therefore are considered as prime candidates for multi-messenger astronomy. \cite{Pad_2014}, on the basis of a joint positional and energetic diagnostic, have suggested a possible association between blazars (BL Lacs) and some neutrino events reported by the IceCube collaboration \citep{ICECube14}. It has also been suggested that blazars could be sites where ultra high energy cosmic rays (UHECRs) are generated \citep[e.g.][]{zhang14}, although a firm association 
has so far been elusive despite the continuous improvement of the available data \citep{pauger14}.

Despite their low space density, compared to all other types of AGN, blazars are detected in high Galactic latitude surveys at all frequencies. 
In some parts of the electromagnetic spectrum (e.g. in the \mw\ and \gr-ray/VHE bands) they are the dominant population \citep[][see also TevCat\footnote{http://tevcat.uchicago.edu}]{GiommiWMAP09,fermi2lac, PlanckERCSC, paper3,Arsioli2015}.  At present about 3,500 such objects are known \citep{bzcat5}.
In the \gr\ band blazars are by far the sources most frequently detected at high Galactic latitudes and therefore are expected to be one of the main 
constituents of the EGB  \citep[e.g.][]{pad1993,giommi2006,dimauro14,ajello2015}.

In a series of papers \citep[][hereafter Paper I, II and III]{paper1, paper2,paper3} we introduced a new scenario for blazars based on light dilution, minimal assumptions on the physical properties of the non-thermal jet emission, and unified schemes. We called this new approach the {\it blazar  simplified view} (BSV). By means of detailed Monte Carlo simulations we showed that the BSV scenario is consistent with the complex observational
properties of blazars as we know them from all the surveys carried out so far in the radio, X-ray, \gr, and VHE bands, solving at the same time a number of long-standing issues. 
%In subsequent papers \citep[][hereafter Paper II and paper III]{paper2,paper3} we extended the Monte Carlo simulations to the \gr\ and VHE bands. 

The purpose of this paper is to go beyond the sensitivity of current surveys and estimate the contribution of blazars to the CXB and EGB, building upon the simulations presented in Papers I, II, and III, and applying the same prescriptions.

As in previous papers we use a  standard $\Lambda$CDM cosmology with $H_0 = 70$ km
s$^{-1}$ Mpc$^{-1}$, $\Omega_m = 0.27$ and $\Omega_\Lambda = 0.73$
\citep{kom11}.

\section{Monte Carlo simulations}\label{ingredients}

In the first two papers of this series we presented the principles on which the BSV is built and mostly concentrated on the statistical properties 
of blazars, like distributions trends, average values of some important parameters, such as the percentage of featureless BL Lacs, the observed peak frequency of the synchrotron component (\nup),  
the relative fraction of FSRQs and BL Lacs, the amount of cosmological evolution, showing that all these are fully consistent with 
the {\it largely different} distributions and values observed in real radio and X-ray surveys. In \paperthree\ we tied our simulation to the absolute numbers of the \fermi-2LAC catalogue 
and we predicted the number of sources in the VHE band taking into account the extragalactic background light (EBL) absorption. In this paper, 
after verifying that our estimates are consistent with the statistical properties of the recently released \fermi-3LAC blazar sample, we extend the simulations below the limits 
of current surveys to calculate the integrated flux from the entire population of blazars of different types in the X-ray and \gr\ bands.
This requires us to explore regions of the (radio) blazar luminosity function not yet observed, and therefore to introduce some additional  
uncertainties arising from the assumptions that we must make on the slope and shape of the luminosity function close to the low luminosity end. 
These uncertainties are however constrained by the fact that here we restrict the parameters of our luminosity function by imposing that the 
predicted soft X-ray (0.3-3.5 keV) and hard X-ray logN-logS are consistent with the available data.  The soft X-ray band, in particular, imposes tight constraints 
on the radio luminosity function since the 5GHz radio flux density of some BL Lacs with X-ray flux above the limits of existing X-ray surveys can be as faint as 
$\sim $1 mJy \citep{stocke91,rector2000,bzcat5}.  

For the reader's convenience we briefly summarise here our Monte Carlo simulations (see \paperone\ for details)
covering the radio through the $\gamma$-ray bands, which are based on a number of ingredients, including: the (radio) blazar luminosity function 
and evolution, a distribution of the Lorentz factor of the electrons and of the Doppler factor, a synchrotron model, an 
accretion disk component, the host galaxy, plus a series of $\gamma$-ray constraints based on observed distributions estimated using 
simultaneous multi-frequency data:  the distribution of Compton dominance, the dependence of the $\gamma$-ray spectral
index on \nup, and that of the $\gamma$-ray flux on radio flux density. 
Sources are classified as BL Lacs, FSRQs, or radio galaxies based on the optical spectrum, as in real surveys. Readers
are referred to \paperone\ and II for full details.

\section{Blazars contribution to extragalactic backgrounds}  
In this section we estimate the contribution to the X-ray and \gr\ cosmic backgrounds by the discrete sources generated in our simulated survey (see Sect. \ref{ingredients}).

The contribution to an apparently diffuse extragalactic background resulting from the integrated  emission from a population of discrete point sources 
can be calculated as follows: $B = \int_{S_{min}}^{S_{max}} S~{dN\over{dS}}~dS$ where $dN\over{dS}$  is the differential logN-logS distribution, $S_{max}$ is the flux of the brightest undetected source, and $S_{min}$ is the flux produced by the least luminous sources in the population at the largest redshift considered.

\subsection{The Cosmic X-ray Background}\label{x background}

The CXB has been convincingly demonstrated to be mostly due to the integrated light generated
by accretion processes onto supermassive black holes at work in AGN \citep[e.g.][and references therein]{ueda2014}. 
This has been proven through two complementary approaches: 1) direct detection in several surveys, down to very faint sensitivities in the soft X-ray band (that is up to $\approx 5 $keV), 
and through relatively shallow surveys in the hard X-ray band (up to $\approx 50$ keV); 2) AGN synthesis models,  which estimate the contribution to the entire CXB combining the luminosity function and cosmological evolution of the sources detected in the soft band with a population of sources that are too heavily obscured (Compton thick AGN) to be detected in soft X-ray surveys \citep[e.g.][]{ueda2014,alexander13}.
These models are quite sophisticated and take into account the latest (radio-quiet) AGN luminosity functions,  cosmological evolution, and parameter correlations. 
However, they ignore the contribution of blazars, which albeit relatively small as we will show later, is not negligible, thus introducing potentially significant uncertainties and biases.

A first attempt at estimating the contribution of blazars to the the X-ray and \gr\ backgrounds was carried out by \cite{giommi2006} using a semi-quantitative method based on the broad-band spectral energy distribution (SED) of blazars of different kind. More recently \cite{ajello2009} made a detailed estimation of the hard-X-ray background based on the Swift-BAT survey in the $15-55$ keV band. 

To calculate the contribution of blazars to the CXB up to its peak near 30 keV, we have extended our Monte Carlo simulations to the hard-X-ray band ($30 - 50$ keV) extrapolating the soft X-ray flux (obtained as described in Paper I) assuming a power law spectral shape with spectral index drawn 
from the the distribution of observed values in the $0.5 - 10$ keV band. The latter was determined from a systematic analysis of all Swift-XRT observations of a sample of more than one hundred bright  blazars (Giommi et al. 2015, in preparation). This was done both for the case where the X-ray emission is dominated by the synchrotron mechanism (High Frequency Peaked Blazars, or HBLs, with steep X-ray spectra) and by inverse Compton emission \citep[Low frequency Peaked Blazars, or LBLs, with flat X-ray spectra; see][for the original definition of LBL and HBL objects]{padovanigiommi95}.

\begin{figure}
\includegraphics[height=6.2cm]{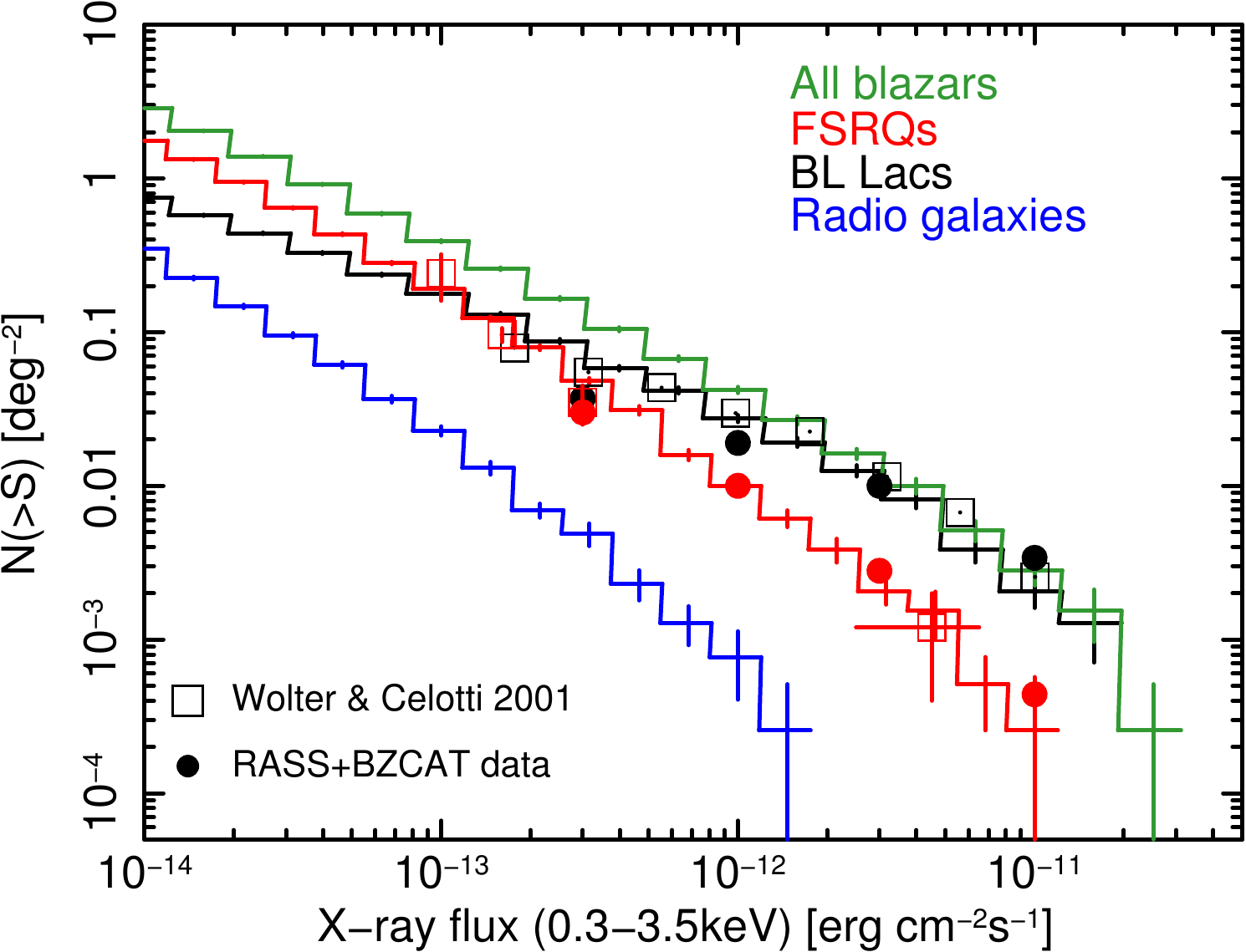}
\caption{The simulated soft X-ray (0.3-3.5 keV) logN-logS of blazars compared to observational data. Filled circles 
refer to our estimates based on ROSAT data, while empty squares are from \protect\cite{WolterCelotti}.}
\label{fig:softx-logns_case5}
\end{figure}

We then calculated the logN-logS for different types of blazars in the soft X-ray ($0.3 - 3.5$ keV) band and compared it with observational data from the extended medium sensitivity survey \citep[EMSS:][]{WolterCelotti} (see Fig. \ref{fig:softx-logns_case5}).
The EMSS covers only less than 4\% of the high Galactic latitude sky, an area that is not large enough to grant a sufficiently large number of FSRQs at fluxes larger than 10$^{-12}$ \ergs. At these high fluxes we estimated the logN-logS using the ROSAT all sky survey \citep[RASS;][]{voges1999,voges2000}, which covers the entire sky (albeit with different sensitivities depending on the exposure), and the latest version of the largest compilation of blazars \citep[BZCAT5\footnote{http://www.asdc.asi.it/bzcat/},][]{bzcat5}. We proceeded as follows: for a given X-ray flux  F($0.3 - 3.5$ keV) we took that part of the RASS survey where the sensitivity was better or equal to F. To avoid complications with the Galactic plane we only considered areas of the sky at $|$b$| > 20^{\circ}$. We then searched for FSRQs in this area by cross-matching the X-ray sources 
with the BZCAT5 list. Since BZCAT5 is not a flux limited nor a complete catalogue, especially at faint fluxes,
the number of FSRQs that match one of the RASS sources divided by the area of sky considered is a lower limit to the density of sources. 
However, at bright X-ray fluxes (say  F $\gsim$ 10$^{-12}$ \ergs) most of the RASS sources have been observed with optical telescopes and identified, and 
therefore the density estimation in this regime is reliable.
Results are summarised in Table \ref{tab:RASSBZCATLogns}  where column 1 gives the X-ray flux F, column 2 gives the number of square degrees where the RASS sensitivity is better than F, columns 3  and 4 give the 
number of FSRQs and BL Lacs with X-ray flux larger than F that match one of the RASS sources within 1 arc minute in the area, while columns 5 and 6 give the surface density of FSRQs and BL Lacs. These density values are plotted in Fig. \ref{tab:RASSBZCATLogns} as filled circles (black for BL Lacs and red for FSRQs\footnote{The ``radio-galaxies'' in the figure are bona-fide blazars misclassified by current classification schemes because their non-thermal 
radiation is not strong enough to dilute the host galaxy component even in the Ca H\&K break region of 
the optical spectrum (see \paperone\ and II for more details).}). The agreement between the data and our simulations 
is good. 

\begin{table*}
%\begin{tiny}
\caption{An estimate of the soft X-ray logN-logS blazar surface density from the RASS and  the Roma-BZCAT catalogue (BZCAT5)}
\begin{tabular}{llrrcc}
 Flux & Area  & No of  & No of & N($>$F) & N($>$F) \\
  erg/cm2/s & sq deg. &FSRQs & BL Lacs& FSRQs& BL Lacs \\
%\multicolumn{5}{l}{~~~~~~~~~~~~~~~~~~~~~~~~~~~~~~Unabsorbed ~~~~~~~~~~~~~~~~~ Absorbed}\\
\hline
%1.e-13  &  560 & 41  &  33 & $>$7.3e-2$\pm$ 1.1e-2 & $>$5.9e-2$\pm$1.0e-2 \\
%3.e-13  &  9300 & 282  & 345  & $>$3.0e-2$\pm$ 1.8e-3 & $>$3.7e-2$\pm$2.0e-3\\
1.e-12&25600&261&  499 & $\gsim$1.0e-2$\pm$6.3e-4 & $\gsim$1.9e-2$\pm$8.7e-4\\
3.e-12&27050&76& 281  & 2.8e-3$\pm$3.2e-4& 1.0e-2$\pm$6.3e-4\\
1.e-11&27100&12&  91 & 4.4e-4$\pm$1.3e-4& 3.4e-3$\pm$3.5e-4\\
\hline
%\multicolumn{3}{l}{\footnotesize $^{*}$Average of 10 runs each simulating 1,000 sources}\\
%\multicolumn{5}{l}{\footnotesize $^{a}$BL Lacs with measurable redshift}
\end{tabular}
%\end{tiny}
\label{tab:RASSBZCATLogns}
\end{table*}

\begin{figure}
\includegraphics[height=6.2cm]{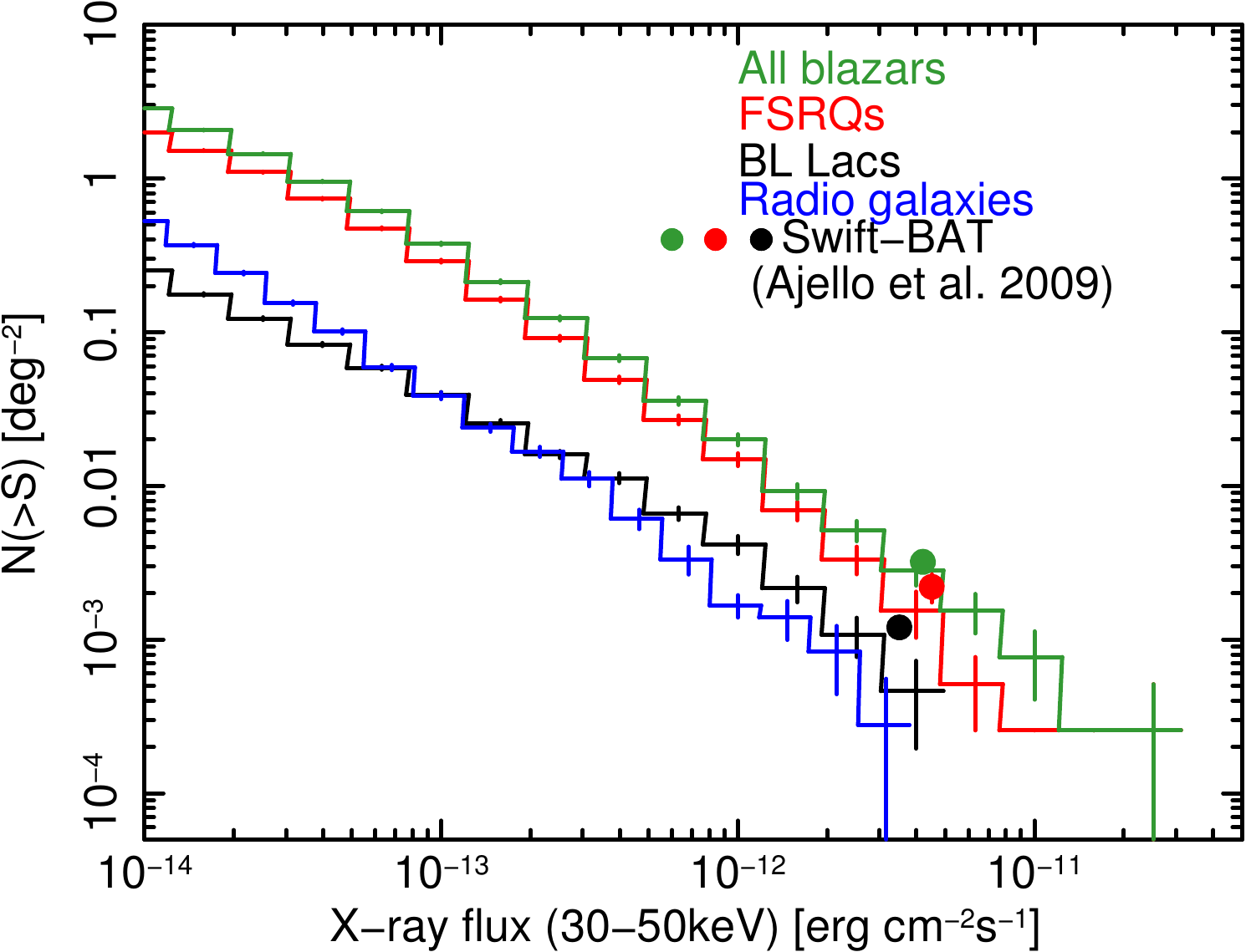}
\caption{The simulated hard X-ray (30-50 keV) logN-logS of blazars is plotted together with  available observed data. The filled circles represents Swift-BAT measurements from \protect\cite{ajello2009}, converted to the 30-50 keV band from the original 15-55 keV band assuming a power law spectrum with spectral index of 1.5 for FSRQs and 2.0 for BL Lacs.}
\label{fig:hardx-logns}
\end{figure}

The simulated $30 - 50$ keV logN-logS is shown in Fig. \ref{fig:hardx-logns}.
Existing surveys in this energy band have been carried out by Swift-BAT and by INTEGRAL \citep[e.g.][]{ajello2009} covering a large fraction of the sky, however  
with sensitivity that is just below $10^{-11}$ \ergs, i.e. orders of magnitude worse than that reached in the soft X-ray band. The observed surface density of 
Blazars is plotted in Fig. \ref{fig:hardx-logns} as filled circles (green for all blazar types, red for FSRQs and  black for BL Lacs). The agreement between the existing data 
and the simulated counts is good, but it is limited to the very brightest end of the logN-logS. Measurements at fainter fluxes will be provided by 
NuSTAR, the first focussing orbiting telescope working in the hard X-ray band, which is capable of detecting serendipitous sources that are up to $\approx$100 times fainter than the sources detected by Swift-BAT or INTEGRAL. However, the field of view of NuSTAR is only a fraction of a degree and it will take some time before sufficient area of sky is covered. At the moment only one blazar has been reported as serendipitously discovered by NuSTAR \citep{alexander13}.

\subsection{Contribution to the  \gr\ (100 MeV - 820 GeV) extragalactic background}\label{gamma background}

\begin{figure*}
\includegraphics[height=12.5cm]{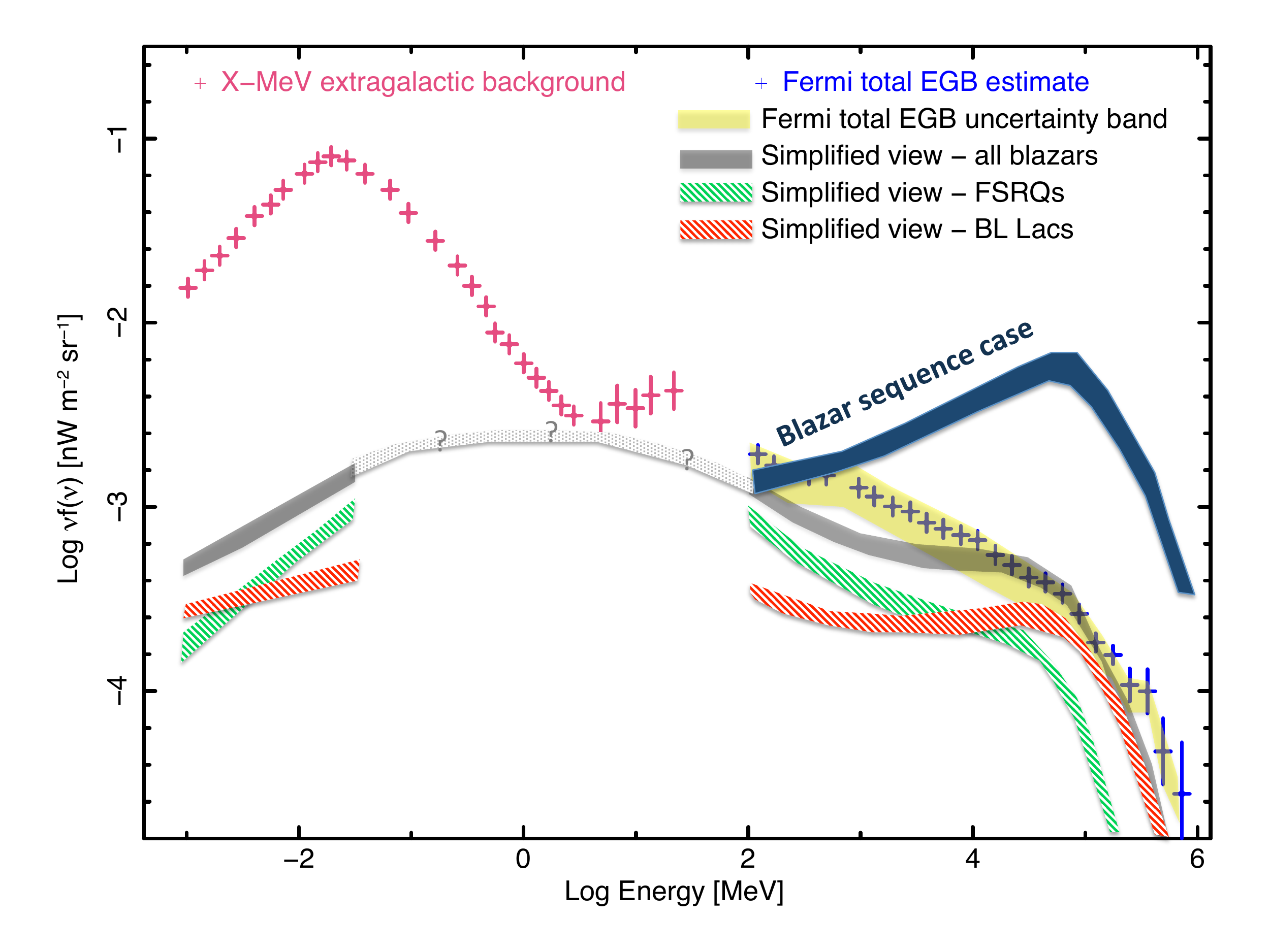}
\caption{The blazar contribution to the X-ray and \gr\ extragalactic backgrounds. All Fermi data are from \protect\cite{fermiegb}. The dark grey, green and 
red bands represent the predictions of the blazar simplified view for all blazar types, FSRQs and BL Lacs, respectively. The uncertainty bands come from different assumptions on the radio luminosity function near the low luminosity end. The light grey area with question marks represents a guess of the possible contribution of blazars between 50 keV and 100 MeV: in this region the SED of blazars is not known sufficiently well to allow a quantitative estimation. The blue band marked "Blazar sequence case" corresponds to the background that is predicted in the case where the peak of the synchrotron emission depends on luminosity like in the "blazar sequence" scenario. The uncertainty band reflects different assumptions on the  luminosity function and a range of possible slopes in the \nup\ vs \nulnu\ correlation, see text for details.}
\label{fig:backgrounds}
\end{figure*}

The most accurate measurement of the Extragalactic Gamma-Ray Background (EGB) available today has been recently reported by the Fermi-LAT team  \citep{fermiegb} and covers the energy range $0.1 - 820$ GeV (see Fig. \ref{fig:backgrounds}).

In our calculation we must take into account the fact that at energies larger than $\sim 50$ GeV absorption by the Extragalactic Background Light (EBL) \citep{stecker92} becomes important. This absorption strongly depends on redshift and on the \gr\ spectral slope ($\Gamma$) in the $0.1-200$ GeV band\footnote{We have assumed a spectral break with $\Delta\Gamma$ = 0.5 at energies $> 200$ GeV.}. We calculate the \gr\ background at energy E  as follows:
$$B(E)=\int_{S_{min}}^{S_{max}}dS\int_{0}^{z_{M}}dz\int_{\Gamma_{m}}^{\Gamma_{M}}Se^{-\tau_{ebl}(E,z)}{dN(E,S,\Gamma)\over{dSdzd\Gamma}}d\Gamma $$
In practice, with our simulated survey data we first build the differential logN-logS in narrow bins of redshift and spectral slopes, then we convert the integrated flux into flux density at energy E using the proper $\Gamma$,  we attenuate the flux for EBL absorption multiplying by $ e^{-\tau_{ebl}(E,z)} $, and then we integrate in redshift and flux. The values of $\tau_{ebl}(E,z)$ were taken from \cite{dominguez11}. 

The results are plotted in Fig. \ref{fig:backgrounds}, where the contribution of blazars of all types is shown as a grey band, representing the uncertainty due to different possible extrapolations of the luminosity function that are consistent with the observed X-ray and \gr\ counts (see Fig. \ref{fig:softx-logns_case5} and \ref{fig:hardx-logns}). 
If we distinguish the contribution of the different blazars subclasses,  Fig. \ref{fig:backgrounds} shows that BL Lacs (red hatched band) are the most important 
contributors in the soft X-ray band and above $\sim 10$ GeV, whereas FSRQs (green hatched band) contribute more flux in the medium-hard X-ray band ($\sim$ 5-50 keV), between 100 MeV and few GeV, and likely also at intermediate energies.

Central to the BSV is the assumption that there is a single mechanism that accelerates particles in FSRQs and BL Lacs and that the maximum energy that the electrons can reach does not depend on luminosity.  In the {\it blazar sequence} scenario \citep[e.g.][]{fossati98,ghis98} instead, the maximum energy of the electrons is a strong function of the source luminosity. We used our Monte Carlo simulations to verify the implications of this dependence on luminosity on the contribution of blazars to the cosmic backgrounds. We have tested the correlation that defines the {\it blazar sequence} by imposing the following dependence of \nup\ on radio luminosity:
Log(\nup ) = a$\times$\lognulnu + b.
   
To take into account of the uncertainties in the correlation we allowed the parameter a to vary from -1.2 to -0.8 and set the constant b in such a way that the resulting dependence of \nup\ on \nulnu\ is consistent with the famous Fig. 12 of  \cite{fossati98}. 
As shown in Fig. \ref{fig:backgrounds} the resulting \gr\ background is comparable to the observed intensity only around 100 MeV, while it rapidly becomes far in excess of it at energies larger then a few GeV.

This excessive background value is due to the fact that in this scenario, HBL sources, which contribute most of the \gr\ background above $\sim 10$ GeV,  are all  constrained in the low power end  of the 
steep radio luminosity function of blazars, on which our simulations are based where the source abundance is largest, thus leading to a very large integrated amount of \gr\ emission.

We note that \cite{ajello2015} in their calculation of the contribution of blazars to the EGB assume that the \gr\ spectral index $\Gamma$  (which is correlated to \nup ) is independent of luminosity  (see their  eq. 12).  With this assumption the agreement between expectations and \fermi-LAT measurement is good. Had they assumed a strong correlation between $\Gamma$ and luminosity as in the {\it blazar sequence}, similarly they would have obtained a largely different expected \gr\ background, likely much larger than observed, as in the case of our simulations.

\subsection{The region between 50 keV and 100 MeV}
In the previous sections we estimated the contribution of blazars to the extragalactic backgrounds in the X-ray ($1-50$ keV) and in the \gr\ (100 MeV -- 820 GeV) bands, where data  from surveys and pointed observations are abundant. Calculating the contribution to the background at intermediate energies is subject to much larger uncertainties as this requires assumptions on the detailed spectral shape of blazars in a region of the electromagnetic spectrum where only very few low-sensitivity observations exists. 
Instead of relying on unconstrained assumptions on the distribution of spectral shapes we decided to make a simple reasonable guess interpolating between the results in 
the X-ray and \gr\ band using a simple geometrical shape. This is shown as a grey dotted band with question marks in Fig. 3. 

\section{Discussion and conclusion}

The  {\it blazar simplified view} is a new paradigm that rests on a minimal set of physical and geometrical assumptions, and on the existence of a universal particle accelerator that produces the same distribution of electron peak energies (see Fig . 4 of \paperone) regardless of luminosity and of blazar type. 

We have shown in previous papers that this scenario reproduces well the statistical properties of blazars in all the energy bands where observational data are available. In this paper we extended our Monte Carlo simulations to calculate the contribution of blazars to the extragalactic cosmic backgrounds in the X-ray and \gr\ bands.  Our results can be summarised as follows:

\begin{enumerate} 
\renewcommand{\theenumi}{(\arabic{enumi})} 
 
\item The contribution of blazars to the 100 MeV to $\approx$ 10 GeV EGB is important but no larger than $\approx$ 50-70\%. 
In this energy band other source types must contribute to the EGB, leaving room also for radio galaxies and star-forming 
galaxies that are also expected to account for a significant fraction of the background \citep[][]{inoue11,dimauro13};

\item Blazars, particularly those of the HBL type, are responsible for 100\% of the EGB above $\approx$ 10 GeV ; 

\item The EGB cutoff at $\sim 250$ GeV \citep{fermiegb} can be fully explained by EBL absorption of the light that would otherwise come from distant ($z >  0.2-0.3$) objects.  
This and the previous point show that our simulations are in good agreement with recent results based on the luminosity function of \fermi-detected blazars \citep{dimauro14,ajello2015}

\item Blazars account for about 4-5\% of the CXB up to $\sim$ 50 keV. 
Assuming that the light grey band with questions marks in Fig. \ref{fig:backgrounds} is a reasonable guess, the contribution of blazars to the 
Extragalactic Cosmic Background between the X-ray and the $\gamma$-ray band becomes about 10\% at 100 keV. Their contribution at larger energies 
increase significantly and steadily until it becomes dominant around a few MeV (see Fig. \ref{fig:backgrounds}).

\item Although blazars account for only  $\sim$ 4-5\% of  the CXB between 1 and 50 keV, this contribution is not negligible and it should be properly taken into 
consideration in CXB synthesis models \citep[see e.g.][and referenced therein]{ueda2014}. This would allow a better estimation of the balance between soft, hard and Compton thick sources in the mix that makes up the CXB. Given that  the contribution to the CXB increases at higher energies the inclusion of 
blazars in CXB synthesis models would have a significant impact on the estimation (or assumptions) of the high-energy cutoff in Seyfert galaxies.  

\item The {\it blazar sequence}  \citep{fossati98,ghis98}, which implies a strong dependence of the synchrotron peak on luminosity is ruled out by the data as it implies a background intensity that is at least ten times the observed value at energies larger than 10 GeV.  

\end{enumerate}

\section*{Acknowledgments}
PP thanks the ASI Science Data Center (ASDC) for the hospitality and
partial financial support for his visits. 
%We acknowledge the use of data and software facilities from the ASDC, managed by the Italian Space Agency
%(ASI). 
%Part of this work is based on bibliographic information obtained 
%from the Astrophysics Data System (ADS). %TBD DO WE NEED TO THANK THE ADS? 

\label{lastpage}
\end{document}